# Photonic Topological Insulators: A Beginner's Introduction

Dia'aaldin J. Bisharat, *Member, IEEE*, Robert J. Davis, *Student Member, IEEE,* Yun Zhou*,* Prabhakar R. Bandaru, and Daniel F. Sievenpiper, *Fellow, IEEE*

*Abstract*— Control and manipulation of electromagnetic waves has reached a new level with the recent understanding of topological states of matter. These metamaterials have the potential to revolutionize many areas in traditional electromagnetic design, from highly robust cavities to small footprint waveguides. Much of the past literature has been on the cutting edge of condensed matter physics, but there is now ample opportunity to explore their usage for practical microwave and optical devices. To assist the beginner, in this Tutorial we give a basic introduction to the essential concepts of topological phenomena in electromagnetic systems, including geometric phases, topological invariants, pseudospin states, and the integer/valley/spin quantum Hall effects. Our focus is on engineered photonic topological insulators (PTIs) in two-dimensional systems. We highlight methods for characterizing such structures and how they result in unique wave-guiding properties. In addition, we provide recipes on how to realize PTIs using photonic crystals and metasurfaces, examine differences between different types of PTIs, and discuss limitations and advantages of some of the existing enabling platforms.

## I. Introduction

MUCH THE same way photonic crystals (PhCs) applied the ideas of solid-state physics to photons [1], i.e. electromagnetic waves, the new field of photonic topological insulators [2] (PTIs) finds its origins in the world of electronic systems. In electronic topological insulators (TIs), electrons propagate along certain directions only on the exterior of the system. This explains part of the name: it is an "insulator" insomuch as it acts like a regular electrical insulator within the bulk of a material. "Topological," on the other hand, comes from the global topology of their energy band structure, since it can be categorized by an integer (the "topological invariant") that does not depend on the fine details of the system (see Fig 1). The occurrence of electrical current on the surface of TIs—and how it responds to changes in energy—is

Submitted for review on 2020-10-28. This work was supported in part by AFOSR contract FA9550-16-1-0093, by ARO contract W911NF-17-1-0453, and DARPA contract W911NF-17-1-0580.
　D. J. Bisharat was with the University of California, San Diego, La Jolla, CA 92093 USA. He is now with the CUNY Graduate Center, New York City, NY 10016. (e-mail: dbisharat@ucsd.edu)
　R. J. Davis, Y. Zhou, P. R. Bandaru, and D. F. Sievenpiper are with the University of California, San Diego, La Jolla, CA 92093 USA. (e-mail: dsievenpiper@eng.ucsd.edu)

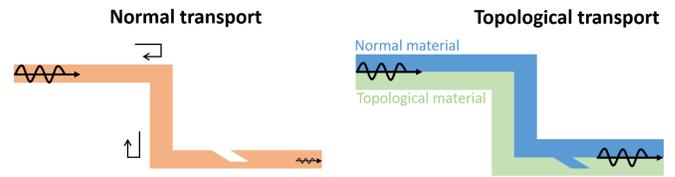

Fig. 1.  Normal vs topologically protected transport. The normal case has backscatter at sharp corners and defects, whereas the topological one does not.

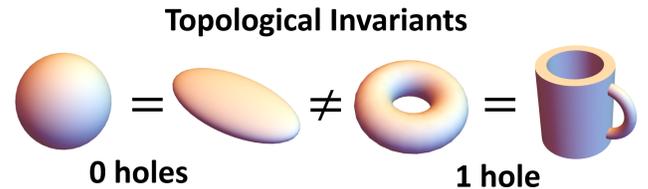

Fig. 2.  Topology concerns quantities that are preserved under continuous deformations of objects. A transformation is "continuous" if it does not cause any sharp cuts or tears in the object. The number of holes in a closed surface is an example of a topological invariant, since a hole cannot be added or removed continuously: a torus can be stretched and pulled into a coffee cup shape, but not a sphere (see above). Electrical conductance in TIs is also determined by a topological invariant, called the Chern number ($C$), where the object is an energy band in the Brillouin zone and the "holes" are determined by the accumulation of the Berry phase.

credited to this topological invariant (see Fig. 2) rather than minor changes to the surface, as in ordinary materials.

TIs found their start in the 1980s with the discovery of the quantum Hall effect (QHE) in a two-dimensional electron gas when subjected to periodic potentials and external magnetic fields [3]-[4]. As in the normal Hall effect, applying a magnetic field causes the electrons to spin in cyclotron orbits, with their frequency being determined by the strength of the $B$ field. When the material is strongly confined to 2D and cooled to very low temperatures, the quantization of the energy of these orbits becomes relevant, with the difference between the allowed energies becoming very large as the field strength increases. When the strength of the $B$ field varies enough to permit or remove an energy level, there will be a sudden jump in the transverse conductivity by an exact multiple of fundamental constants. Hence, the QHE shows that conductivity is fundamentally *discrete* [5]. Importantly, it was found [6] that this discrete behavior could be explained by a special phase (called the geometric, or Berry phase, detailed in the next section) that each electron accumulates as it orbits in cyclotron motion in reciprocal ($k$) space.

How does topology relate to this? As it turns out, the discrete nature of the conductance is highly robust to deformations to the bulk of the material, and it can be shown that the added geometric phase responsible for the quantization is tied directly to the mathematical framework of topological invariants (see Fig. 2) [3], [6]-[7]. This has some important consequences: it gives us a simple means to classify materials (i.e., bandgap materials) by calculating their topological invariant (which is a property of the bulk material), and it results in the technologically useful effects that topological insulators offer.

Materials that have an invariant of zero are "trivial," and act the same as an ordinary material. If the invariant is nonzero, however, then the effects of the geometric phase become relevant, and "non-trivial" effects can be observed. One of the most startling effects is what happens at the edge between a non-trivial material and a trivial material (or another non-trivial material with a different invariant), where a highly robust transport mode can exist [5]. These special modes, called edge modes, exist *within* the bandgap of the non-trivial material, and can be explained by the sudden change in the invariant across the boundary (e.g., going from 1 to 0). Even more remarkable is that electrons moving along these boundaries must do so in one direction only, with no possibility of scattering back in the other direction (illustrated in Fig. 1) [8]. These edge modes are the corollary of the quantization of conductivity in the QHE.

In repeated experiments these edge states are observed regardless of the impurities in different material samples [9]. Since the invariant is resistant to a wide range of distortions to the material, the edge states are said to be topologically "protected," guaranteed to exist so long as the invariant stays the same [5]. This is of technological importance due to the potential to reduce power consumption by eliminating sources of loss, as well as simplify manufacturing by increasing defect tolerances. These discoveries led to the Nobel Prize in Physics being awarded to Thouless, Haldane, and Kosterlitz in 2016.

These systems with topological behavior are a consequence of the wave nature of the electrons, not specifically their quantum interactions [10]. As a result, it is possible to construct classical wave systems with analogous properties to their electronic counterparts. This opens the door to a vast range of theoretical proposals and experimental demonstrations. Replacing the electron with a photon (along with a reinterpretation of some quantities) we arrive at PTIs, which demonstrate many of the same features of TIs [2] and are the primary subject of this tutorial.

In this Tutorial, we outline the basic concept of the geometric phase and extend it to periodic systems in which topological properties emerge. We focus on 2D PhCs and showcase the physical implications of Chern numbers and topological transitions that can arise in such systems. In addition, we discuss the formation of degeneracy (Dirac) points in PhCs and then the various mechanisms to introduce topological phases that make different types of PTIs. Finally, we discuss recent developments and future perspectives on this emerging field.

## II. GEOMETRIC PHASE

The key idea behind topological effects in all areas is the geometric phase, a universal concept that emerges when a parameter describing a system is gradually varied in a closed cycle [11]. This phase was first proposed in 1956 by Pancharatnam for the propagation of light through a sequence of polarizers [12] and was later generalized by Berry for quantum mechanics [13]. Many phenomena in physics can be attributed to the geometric phase, from the mechanical Foucault pendulums [14] to the polarization in helical waveguides [15].

Any wave possesses an amplitude (call it $E_0$) and an ordinary phase ($\phi$) at a given position and time, $E(r,t) = E_0(r)e^{j\phi t}$. When the values of $r$ and $t$ are slowly changed from $r_0$ and $t_0$ to distinct intermediate values $r_1$, $t_1$, then smoothly changed back to $r_0, t_0$, we would intuitively expect that the initial value of $E(r,t)$ would be exactly the same as the final value. However, there are some physically important cases where this intuition fails, as in the case shown in Fig. 3.

If we take a tangent vector (the red arrow) and slide it along a path on a sphere (the black arrows), returning it to the starting position, the arrow will no longer be pointing the same direction. Hence, the starting value no longer matches the final value. Upon its return to the north pole, the orientation of the vector is rotated by an angle $\phi$, which in this case is equal to $\pi/2$. Note that this is only the case because the path is a closed loop: if the tangent retraced the same path to return to the start (enclosing no area), the orientations would match and $\phi$ would be zero. In the closed path case, we can consider the angle an added phase, the geometric phase, that causes the initial and final values to differ. This phase is called

**Side note:** Although the field of PTIs originated from the electronic version, there are some fundamental differences between the two. The most significant is that photons are bosons, whereas electrons are fermions. This difference manifests itself in the ways that different symmetries (like time reversal, written as an operator $T$) can change how a system behaves.

Specifically, time reversal for fermions has the relationship $T^2 = -1$, whereas for bosons it is $T^2 = +1$. A more practical concern is that absorption in the medium can be an issue in photonics. Nevertheless, many of the most technologically relevant features of TIs (including backscatter immunity) can still be found in photonic systems, so long as care is made to distinguish the circumstances in which they can exist.

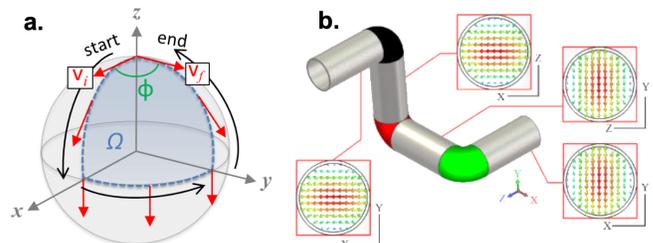

Fig. 3. a. Geometric phase from parallel transport. b. The polarization (shown as the vector E field inset at different locations) in a bent circular waveguide shows that as the propagation path is varied, then returned to its initial state, there can be a phase shift to the polarization state, which is due to the geometric phase. b. is © 2017 IEEE. Reprinted, with permission, from [16].

"geometric" because it corresponds to a geometric area (Ω, shaded in blue) of parameter space that the path encloses.

If we think of the sphere in Fig. 3a as the sphere of constant wavevector $\boldsymbol{k}$, and the vector the $E$ field, this parallel transport explains the change in polarization in helical and bent waveguides (Fig. 3b) [16]. In the example of a bent waveguide, if polarized light gradually changes direction from $z$ to $-x$, then from $-x$ to $y$, and finally back to $z$, the wave will pick up a geometric phase that is added to the complex exponential form of the $E$ field. The "path" in this case is the path traced over the $\boldsymbol{k}$-sphere as the wavevector, corresponding to the direction, is changed. In this case, the extra phase (in the form of polarization direction) can be attributed to the electric field always being perpendicular to the direction of propagation, so as the direction of propagation is changed the polarization must necessarily be altered, despite the propagation direction eventually returning to the starting value. Note that this effect is possible only due to the existence of two degenerate modes in the waveguide (from the circular symmetry).

## III. BERRY PHASE, BERRY CONNECTION, AND BERRY CURVATURE

A geometric phase can emerge due to the gradual variation of a state in many types of parameter spaces, including the momentum space of a periodic system, like those of a PhC [17]. For any path that traverses an allowed band of a periodic system and does not intersect with any other band, the wavevector $\boldsymbol{k}$ (Bloch momentum) varies in closed loops due to the lattice periodicity, where $\boldsymbol{k}_{-\pi} \equiv \boldsymbol{k}_\pi$. In a 2D crystal, $\boldsymbol{k}$ traverses the surface of a torus geometry which bounds the entire Brillouin zone (BZ), see Fig. 4. Many of the most important topological properties appear in such systems and are a simple platform to understand how they emerge.

The literature on topological insulators employs a great deal of terminology, most of which merely refer to a few mathematical constructions that assist in characterizing when a topological invariant is non-trivial. Fortunately, most of these constructions have parallels to standard electromagnetic theory and provide straightforward methods to numerically calculate topological features of real systems.

Consider a lattice described by a general eigenvalue problem in momentum space,
$$H(\boldsymbol{k}) \cdot \boldsymbol{\psi}_n(\boldsymbol{k}) = \lambda_n(\boldsymbol{k})\psi_n(\boldsymbol{k}) \quad (1)$$
where $\lambda_n(\boldsymbol{k})$ is the eigen energy and $\boldsymbol{\psi}_n(\boldsymbol{k})$ is the normalized eigen wavefunction of $H(\boldsymbol{k})$ (often called the "Hamiltonian" in the literature) at each $\boldsymbol{k}$ for the $n^{th}$ band, which can be determined via Bloch's theorem. In the following, we will make use of the shorthand notation of inner product
$$\langle A(\boldsymbol{r})|B(\boldsymbol{r})\rangle \equiv \int A(\boldsymbol{r})^\dagger \cdot B(\boldsymbol{r}) d\boldsymbol{r} \quad (2)$$
to refer to integration of two vector functions $\boldsymbol{A}$ and $\boldsymbol{B}$ over a variable $\boldsymbol{r}$, with † denoting Hermitian conjugation. Hence, normalized in this case means $\langle \boldsymbol{\psi}_n(\boldsymbol{k})|\boldsymbol{\psi}_n(\boldsymbol{k})\rangle = 1$. Gradually changing the $\boldsymbol{k}$ along a given energy band will cause a phase accumulation associated with the slow evolution of $\boldsymbol{\psi}_n(\boldsymbol{k})$. Under most cases when $\boldsymbol{k}$ returns back to where it started this accumulation will result in zero total phase, but, like in the examples of Section II, special cases can arise where a non-zero geometric phase is added. In PTI literature the geometric phase is referred to by the name **Berry phase**, specifically to recall Berry's formulation in quantum mechanics [13].

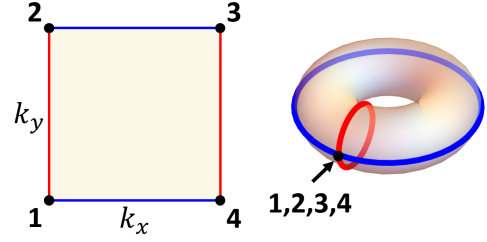

Fig. 4. The Brillouin zone can be considered as a torus by taking each periodic boundary (red and blue in the figure) and connecting them together.

To calculate the total Berry phase, we need a means to add up the phase contributions from each small change to the wavefunction. The phase shift between two $\boldsymbol{\psi}_n$ states infinitesimally separated by $\Delta \boldsymbol{k}$ can be represented by their inner product [16], expanded as a low-order Taylor series as
$$\langle \boldsymbol{\psi}_n(\boldsymbol{k})|\boldsymbol{\psi}_n(\boldsymbol{k}+\Delta\boldsymbol{k})\rangle \approx 1 + \Delta\boldsymbol{k}\langle \boldsymbol{\psi}_n(\boldsymbol{k})|\nabla_{\boldsymbol{k}}|\boldsymbol{\psi}_n(\boldsymbol{k})\rangle$$
$$= \exp[-i\Delta\boldsymbol{k} \cdot \boldsymbol{A}_n(\boldsymbol{k})]. \quad (3)$$
Here, we can see that $\Delta\boldsymbol{k} \cdot \boldsymbol{A}_n(\boldsymbol{k})$ is the phase shift over $\Delta\boldsymbol{k}$, and $\boldsymbol{A}_n(\boldsymbol{k})$ is the rate of change of the phase shift. $\boldsymbol{A}_n(\boldsymbol{k})$ is called the **Berry connection**, or **Berry vector potential**,
$$\boldsymbol{A}_n(\boldsymbol{k}) = i\langle \boldsymbol{\psi}_n(\boldsymbol{k})|\nabla_{\boldsymbol{k}}|\boldsymbol{\psi}_n(\boldsymbol{k})\rangle. \quad (4)$$

> **Side note:** The notation $\langle \alpha|\beta \rangle$ represents the inner product of the wavefunctions $\alpha$ and $\beta$, whereas $\langle \alpha|\nabla_{\boldsymbol{k}}|\beta \rangle$ denotes the inner product of $\alpha$ and $\nabla_{\boldsymbol{k}}\beta$.

Therefore, the **Berry phase** for the $n^{th}$ band is defined as the integral of $\boldsymbol{A}_n(\boldsymbol{k})$ over some closed path $l$ in $\boldsymbol{k}$-space,
$$\phi_n = \oint_l d\boldsymbol{k} \cdot \boldsymbol{A}_n(\boldsymbol{k}). \quad (5)$$
The path $l$ is simply a smooth curve of values over the BZ, such as the blue and red lines shown on the right half of Fig. 4. If we know what a given wavefunction looks like in the Brillouin zone, we could use the above to calculate the Berry phase for that path. However, there is a catch: the Berry connection $\boldsymbol{A}_n(\boldsymbol{k})$ is not uniquely defined. If a phase change $\zeta(\boldsymbol{k})$ is added to the eigen wavefunction $\boldsymbol{\psi}_n(\boldsymbol{k})$, where $\zeta(\boldsymbol{k})$ is a periodic function with $\zeta(\boldsymbol{k}_{\text{end}}) = \zeta(\boldsymbol{k}_{\text{begin}}) + 2m\pi$, the new wavefunction $e^{i\zeta(\boldsymbol{k})}\boldsymbol{\psi}_n(\boldsymbol{k})$ is still an eigen wavefunction to $H(\boldsymbol{k})$. The Berry connection is then transformed as $\boldsymbol{A}_n(\boldsymbol{k}) \to \boldsymbol{A}_n(\boldsymbol{k}) - \frac{\partial}{\partial \boldsymbol{k}}\zeta(\boldsymbol{k})$, where it changes its formula with different choice of $\zeta(\boldsymbol{k})$.

The Berry phase, on the other hand, is invariant modulo $2\pi$,
$$\oint_l d\boldsymbol{k} \cdot \boldsymbol{A}_n(\boldsymbol{k}) \to \oint_l d\boldsymbol{k} \cdot \boldsymbol{A}_n(\boldsymbol{k}) - \oint_l \frac{\partial}{\partial \boldsymbol{k}}\zeta(\boldsymbol{k})d\boldsymbol{k}$$
$$\to \oint_l d\boldsymbol{k} \cdot \boldsymbol{A}_n(\boldsymbol{k}) - 2m\pi. \quad (6)$$
This can also be understood qualitatively. As the wavevector $\boldsymbol{k}$ slowly travels around the loop of a band, the wavefunction $\boldsymbol{\psi}_n(\boldsymbol{k})$ eventually returns to where it starts and picks up a phase of a multiple of $2\pi$, with most systems picking up zero [11]. Since the Berry connection depends on how we setup the calculation, yet know that the Berry phase should not, it is helpful (especially for numerical purposes) to define a quantity that will be invariant to any arbitrary phase $\zeta(\boldsymbol{k})$ that we may add.

The **Berry curvature**, or **Berry flux**, a quantity that is invariant under such transformation, can be constructed by taking the curl the Berry connection,

$$\boldsymbol{\Omega}_n(\boldsymbol{k}) = \nabla_{\boldsymbol{k}} \times \boldsymbol{A}_n(\boldsymbol{k}). \quad (7)$$

Then, using Stokes' theorem, the Berry phase can be rewritten as the integral of the Berry curvature,

$$\phi_n = \int_S d^2\boldsymbol{k} \cdot \boldsymbol{\Omega}_n(\boldsymbol{k}), \quad (8)$$

where the integration is over the surface bounded by the path $l$ [16].

## IV. Topology in 2D Photonic Crystals

The previous section dealt with the general concepts of geometric phase in periodic media, regardless of physical setting (i.e., electronic, photonic, etc.). To make this explicit, here we show how this theory can be specialized for 2D electromagnetic systems. For electromagnetic waves, the eigenvalue problem space is described by the macroscopic Maxwell equations. For non-bianisotropic materials in 2D the magnetic field can be eliminated—for simplicity when treating TM modes, given by the $E_z$ field alone—and Maxwell's equations can be recast in the compact form

$$\nabla_{\boldsymbol{r}} \times [\mu^{-1}(\boldsymbol{r})\nabla_{\boldsymbol{r}} \times E_z(\boldsymbol{r})] = \omega^2 \epsilon(\boldsymbol{r}) E_z(\boldsymbol{r}), \quad (9)$$

where $\omega$ is the angular frequency, $E_z(\boldsymbol{r})$ is the $z$ component of the electric field (hereafter we will drop the $z$ subscript), and $\mu(\boldsymbol{r})$ and $\epsilon(\boldsymbol{r})$ are the magnetic permeability and dielectric permittivity tensors, respectively. Note that we are ignoring dispersive effects for now, but further analysis shows this is valid in many cases [16]. By applying Bloch's theorem to the above equation, the eigen wavefunction can be obtained in the form of eigen electric field $E_{n,\boldsymbol{k}}(\boldsymbol{r})$, assuming a periodicity of the material parameters [1].

Since the eigenvalue problem involves the dielectric permittivity $\epsilon(\boldsymbol{r})$ on the right-hand side of Eq. (9), the inner product of two eigen wavefunctions $E_{n,\boldsymbol{k}_1}(\boldsymbol{r})$ and $E_{n,\boldsymbol{k}_2}(\boldsymbol{r})$ can be written as

$$\langle E_{n,\boldsymbol{k}_1}(\boldsymbol{r})|E_{n,\boldsymbol{k}_2}(\boldsymbol{r})\rangle = \int d^2\boldsymbol{r} E^*_{n,\boldsymbol{k}_1}(\boldsymbol{r})\epsilon(\boldsymbol{r})E_{n,\boldsymbol{k}_2}(\boldsymbol{r}), \quad (10)$$

where ∗ denotes complex conjugation. The Berry connection then takes the form

$$\boldsymbol{A}_n(\boldsymbol{k}) = i\langle E_{n,\boldsymbol{k}}|\nabla_{\boldsymbol{k}}|E_{n,\boldsymbol{k}}\rangle$$
$$= i\int d^2\boldsymbol{r} E^*_{n,\boldsymbol{k}}(\boldsymbol{r})\epsilon(\boldsymbol{r})\nabla_{\boldsymbol{k}} E_{n,\boldsymbol{k}}(\boldsymbol{r}). \quad (11)$$

As in the general case, $E_{n,\boldsymbol{k}}$ is normalized such that $\langle E_{n,\boldsymbol{k}}|E_{n,\boldsymbol{k}}\rangle = 1$.

Subsequently, the Berry curvature and Berry phase can be written as discussed above.

### A. Chern Number

With the aid of the Berry curvature, we can calculate the Berry phase that a given electromagnetic mode may acquire for a 2D PhC lattice. As mentioned in Section I, the relationship to topology comes in the form of an invariant tying a non-zero Berry phase to the edge modes. For electromagnetic systems, this invariant is called the Chern number, after Chinese-American mathematician Shiing-Shen Chern.

The Chern number always takes an integer value. When the Chern number is non-zero, the 2D photonic system is said to be **topologically nontrivial**. The **Chern number** of the $n^{th}$ band of a 2D lattice is simply the Berry phase over the full Brillouin zone,

$$C_n = \frac{1}{2\pi}\int_{BZ} d^2\boldsymbol{k}\Omega_n(k_x, k_y), \quad (12)$$

where in 2D the Berry curvature only has two terms

$$\Omega_n(k_x, k_y) = \frac{\partial A^n_{k_y}}{\partial k_x} - \frac{\partial A^n_{k_x}}{\partial k_y}, \quad (13)$$

where $A^n$ is the Berry connection for the $n^{th}$ mode,

$$A^n_{k_x} = i\int d^2\boldsymbol{r}\, E^*_{n,\boldsymbol{k}}(\boldsymbol{r})\epsilon(\boldsymbol{r})\frac{\partial E_{n,\boldsymbol{k}}(\boldsymbol{r})}{\partial k_x}, \quad (14)$$

$$A^n_{k_y} = i\int d^2\boldsymbol{r}\, E^*_{n,\boldsymbol{k}}(\boldsymbol{r})\epsilon(\boldsymbol{r})\frac{\partial E_{n,\boldsymbol{k}}(\boldsymbol{r})}{\partial k_y}. \quad (15)$$

When calculated for an arbitrary polarized band over the whole BZ, the Chern number as expressed in $C_n$ takes a non-zero value only when time reversal symmetry is broken for the lattice [2]. The most common case when this happens is if a magnetic field is applied (the Faraday and Kerr effects). In such cases, the system is often called a Chern insulator, or Chern PTI.

However, there are a few ways to observe topological effects even when time reversal symmetry is retained [4]. Such systems are *still reciprocal* (i.e., they cannot form true isolators), but they can still display immunity to certain types of backscatter and can act as robust polarization filters [18].

> **Side note:** Why is the Chern number always an integer? A simple explanation comes from comparing the equations with those of the magnetic field. The Berry curvature has the same form as the magnetic field, where the Berry connection takes the place of the magnetic vector potential (hence the term "Berry vector potential"; likewise, the Berry phase can be thought of as a magnetic flux):
> $$\boldsymbol{\Omega}_n = \nabla_{\boldsymbol{k}} \times \boldsymbol{A}_n \Leftrightarrow \boldsymbol{B} = \nabla_{\boldsymbol{r}} \times \boldsymbol{A} \quad (S1a)$$
> Since the Chern number is just the integral of the Berry curvature, this is the $\boldsymbol{k}$-space version of integrating the magnetic field,
> $$C_n = \int \boldsymbol{\Omega}_n \cdot d\boldsymbol{k} \Leftrightarrow \int \boldsymbol{B} \cdot d\boldsymbol{r} \quad (S1b)$$
> From Gauss' law for $\boldsymbol{B}$ fields, we know that doing so will always give zero, unless there exists a magnetic monopole in the integration area. In such a case, the integral will give an integer multiple of monopole charges. In contrast to the $\boldsymbol{B}$ field, the Berry phase *can* contribute "monopoles" to the Berry curvature, the number of which is the Chern number [16]. Hence, the Chern number must be an integer. Further proofs can be found in [2], [7].

One popular version of time reversal symmetric PTIs is the "valley" PTI, which associates different directions (Γ to K and Γ to K') in $\boldsymbol{k}$-space with "valley Chern numbers." To calculate the valley Chern number, $C_v$, the integral in Eq. (12) is simply performed over only one half the BZ, such that two ordinarily identical high-symmetry points (K and K') are separated. This gives two different values of $C_v$, one for each half of the BZ. Added together they will equal the normal Chern number (zero for reciprocal systems) but considered separately they can be non-zero under special cases [19].

The other major reciprocal PTI is the "spin" PTI, which associates the handedness of a circularly polarized mode (or other combinations of modes) with a "spin Chern number," $C_s$ [20]. In general, these spins/polarizations are constructed by a superposition of two or more eigen fields from multiple bands at the same frequency. Each polarization (right-handed and left-handed) corresponds to its own value of $C_s$ [20]-[21]. As such, the electric field in the inner product definition must be replaced with the field associated with a given polarization.

> **Side note:** There are serval ways of defining a spin Chern number. The different formulations depend on how the spins are rigorously related to a true topological invariant of the physical system. In general, the invariant of a spin PTI is frequently referred to as a $Z_2$ topological invariant [22], distinct from the ordinary Chern number, but in this Tutorial it sufficient to consider it as a subtype of the standard Chern number $C_n$, albeit specialized to a given spin definition.

*B. Numerical Calculation of the Chern Number*

In calculations of the Chern number for simulations or experiments, we need to discretize the continuous 2D BZ into a lattice, as shown in Fig. 5. The shown discretization is for a square BZ (for square lattices), but the same methods will work on triangular lattices, where the usual hexagonal BZ is shifted to form a rhombus [23],[24]. The Chern number can be written as [23],[24]

$$C_n = \frac{1}{2\pi} \sum_{k_x,k_y} \Omega_n(k_x, k_y) \Delta k_x \Delta k_y, \quad (16)$$

where

$$\Omega_n(k_x, k_y)\Delta k_x \Delta k_y = \left(A^n_{k_y}(k_x + \Delta k_x, k_y) - A^n_{k_y}(k_x, k_y)\right)\Delta k_y - \left(A^n_{k_x}(k_x, k_y + \Delta k_y) - A^n_{k_x}(k_x, k_y)\right)\Delta k_x. \quad (17)$$

Since $\Delta k_x$ is small,
$$A^n_{k_x}(k_x, k_y)\Delta k_x \approx i\ln\langle E_n(k_x, k_y) | E_n(k_x + \Delta k_x, k_y)\rangle. \quad (18)$$

If we number the four vertices of a small cell 1, 2, 3, and 4 in a clockwise direction, as shown in Fig. 5, the integral of the Berry curvature over the grid can then be written as

$$\Omega_n(k_x, k_y)\Delta k = -i(\ln\langle E_{n,1}|E_{n,2}\rangle + \ln\langle E_{n,2}|E_{n,3}\rangle + \ln\langle E_{n,3}|E_{n,4}\rangle + \ln\langle E_{n,4}|E_{n,1}\rangle)$$
$$= -i\ln(\langle E_{n,1}|E_{n,2}\rangle\langle E_{n,2}|E_{n,3}\rangle\langle E_{n,3}|E_{n,4}\rangle\langle E_{n,4}|E_{n,1}\rangle) \quad (19)$$

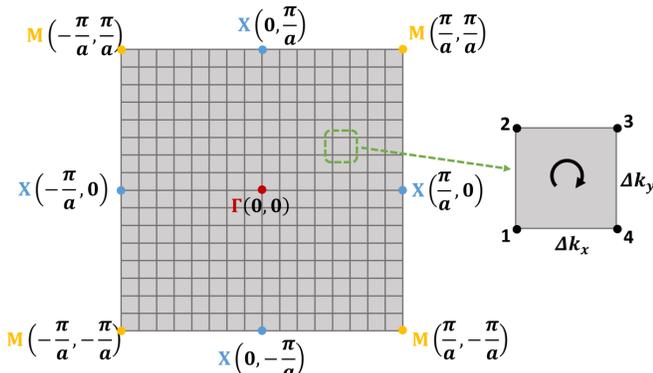

Fig. 5. Discretization of a square 2D BZ in the increment of $\Delta k_x$ in $x$ direction and $\Delta k_y$ in the $y$ direction. A hexagonal BZ can be discretized into a parallelogram grid in a similar manner.

Here,

$$\langle E_{n,p}|E_{n,q}\rangle = \sum_{w,m} E^*_{n,p}(w,m)\epsilon(w,m)E_{n,q}\Delta s \quad (20)$$

where $p, q = \{1, 2, 3, 4\}$ denotes the four vertices, $(w, m)$ indicates the $(w, m)^{th}$ discretized cell in the real space, $\Delta k = \Delta k_x \Delta k_y$, and $\Delta s$ is the area of the discretized lattice in the real space. Eq. (19) shows that the integral of Berry curvature over each small cell in the BZ can be obtained by taking the inner products of the eigen electric fields at adjacent vertices in a clockwise order, as illustrated by the inset in Fig. 5.

Substituting Eq. (19) into the summation in Eq. (16), we get a discrete approximation to the Chern number. It can be shown that this approximation converges to the (continuous variable) Chern number in the limit $\Delta k_{x(y)} \to 0$ [25]. Fortunately, it also rapidly converges; often as coarse a grid as 24 x 24 cells is enough for accurate determination of the Chern number [23],[25]. The spin Chern number can be computed by separating the two distinct spin eigenmodes (generally polarization-based) and performing the Chern number calculation on each [20]. For non-trivial spin PTIs, this will result in two identical values, each being the negative of the other [23]-[22]. For valley Chern number calculations, only half of the BZ is integrated in the above equation to account for the contribution of a finite region in momentum space that correspond to specific high-symmetry points in the BZ [23].

> **Side note:** Computing the Berry curvature and topological invariant for degenerate modes is slightly different than that for the other types, since they involve a combination of modes. To account for multiple modes, it is necessary to change the representation of the Berry connection to a matrix where each element is the Berry connection between a pair of modes [23]. For the common case of a single degenerate pair the matrix takes the form
> 
> $$S_{kk'} = \begin{bmatrix} \langle E^1_k|E^1_{k'}\rangle & \langle E^1_k|E^2_{k'}\rangle \\ \langle E^2_k|E^1_{k'}\rangle & \langle E^2_k|E^2_{k'}\rangle \end{bmatrix}, \quad (S2a)$$
> 
> where the superscripts correspond to the band index. With this formulation, the Berry curvature for a given cell is given by
> 
> $$\Omega(k_x, k_y)\Delta k_x \Delta k_y = -i(\ln(\det(S_{k_1k_2}S_{k_2k_3}S_{k_3k_4}S_{k_4k_1}))). \quad (S2b)$$
> 
> The above can be generalized to as many degenerate bands as necessary [23],[25].

From a practical perspective, the Chern number gives a straightforward means of checking whether a given system has edge states, and therefore if it will be robust to various forms of disorder. The authors have provided a collection of general-purpose MATLAB® functions that perform the various steps, available via a public repository [26]. It is worth noting that these numerical methods are not the only option to determine the nontriviality of a system, with another powerful technique being the Green's function approach (which also simplifies the analysis of degenerated bands) [27]. In the proceeding sections we will show example calculations of each PTI type (Chern, valley, and spin) using this code, with the eigenmode data simulated via Ansys HFSS.

The Chern number describes the topology of a band and characterizes the most fascinating and technologically relevant phenomena: topologically "protected" edge states. These edge states appear at the interface between two structures with unequal Chern numbers.

Unlike traditional photonic waveguides, with a *trivial* edge state between two ordinary insulators (with Chern number of zero), the *nontrivial* edge waveguide formed by these two topologically inequivalent structures (at least one structure is of nonzero Chern number), would be immune to defects and backscattering. This is because when the two domains with different topological invariant are connected directly to form an interface, a **topological phase transition** must happen at the interface [2].

> **Side note:** A **topological phase transition** is the procedure where a system changes its topological invariant. Systems with different topological invariants cannot change into each other without a phase transition. In periodic systems, this occurs when a bandgap closes, marked by a change in Chern number. In PhCs, a topological transition is usually induced by changes in symmetry or geometry of the PhC unit cell.

Essentially, the differing topologies mean that the respective bands in each bandgap material cannot be continuously transformed into one another. Transforming one into the other will require the frequency gap to close at the interface, then reopen on the other side. This phase transition gives rise to the gapless edge states at the interface. To accommodate the jump in the Chern number's integer value, *e.g.* from 1 to 0 or 1 to -1, etc., the number of gapless edge modes turns out to be the difference of the Chern numbers across the interface [28]. This is known as **the bulk-edge correspondence** [2].

These gapless modes are tied to the bulk Chern numbers, so they are robust and must always exist, regardless of the specific shape of the boundary (unlike conventional waveguides) [2]. It is worth stressing that these modes are distinct from those found in standard PhC waveguides (which can also possess high robustness [29]), with the primary difference being their immunity to certain forms of scattering being a *global* property of the bulk, rather than any specific arrangement of PhC cells.

## V. CHERN PTIS

In general, PhCs and other periodic structures have zero Chern number [2]. To engineer one, we need to focus on two key steps: 1. Find a degenerate point between bands, and 2. Break a symmetry that opens a bandgap near that point. This section will detail how the simplest type of PTI, the Chern PTI, is constructed and demonstrate the exciting features it has for practical designs. This type of structure is a direct emulation of the quantum Hall effect discussed previously [30].

The first step, finding degeneracies, relates to the abrupt nature of the Chern number: a material can only change its Chern number (a topological phase transition) when two or more photonic bands are degenerate at a point. This is part of the reason for the robustness of edge modes, as any small change to the structure that keeps the bandgap open in the bulk will not affect the mode. Finding a degeneracy in PhCs is common, but the second step, opening a bandgap via a broken symmetry, places some restrictions on the degeneracies that are useful for making a PTI [2].

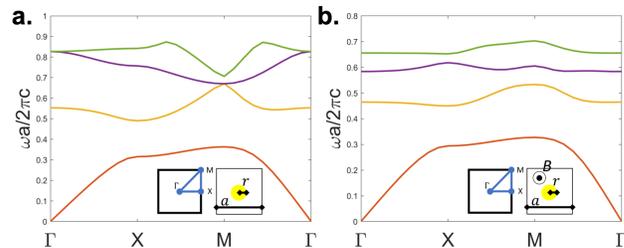

Fig. 6. Band diagram for the square lattice of a. un-magnetized, b. magnetized YIG rods in air. Inset left: BZ path, right: unit cell with $r = 0.11a$.

The simplest type of degenerate point for PTIs is a linear

> **Side note:** The most useful symmetries for EM systems are
> 1. Time Reversal Symmetry (**TRS**): $t \to -t$, broken by gyrotropic materials with applied E/H fields
> 2. Spatial Inversion Symmetry (**SIS**): $r \to -r$, broken by altering the geometry of the material (e.g., removing a slice from cylinder or changing circles to triangles) crossing of two bands, often called a "Dirac cone" in the literature [4]. Such a crossing can be made via a PhC in a honeycomb lattice, where the degenerate point will always occur at the K(K') high symmetry point in the BZ [1]. To obtain a non-trivial PhC it is, however, not necessary to form a linear-type degeneracy, as any other type (e.g., quadradic [31] or accidental [32]) will also work.

To see how this works for a real device, we will use the example of Wang et al. [33], which was later developed into the first experimental demonstration of a Chern PTI [34]. First, to create the initial degenerate point, we select a square lattice of circular rods (chosen to be made of yttrium-iron-garnet, YIG, for reasons soon explained), and tune the geometry to find a quadratic crossing of the 2nd and 3rd bands at the M point, shown in Fig. 6a. Note that there is also a degeneracy at the Γ point for the 3rd and 4th bands, but we will focus on the quadratic M point here.

Now that we have our degeneracy, we must break a symmetry that opens a complete bandgap near it. The chief symmetries present in the system are time reversal symmetry (**TRS**, where running time backwards does not affect the response) and spatial inversion symmetry (**SIS**, where flipping the coordinate axes maintains the shape and orientation of the unit cell).

Breaking either will induce a bandgap, but only breaking TRS will cause a non-zero accumulation of Berry phase over the whole BZ, and so will result in the desired edge modes [24]. Following Wang et al.'s approach, TRS can be broken by applying a static magnetic field perpendicular to the 2D plane. Doing so induces an anisotropy to the magnetic permeability of the YIG with form

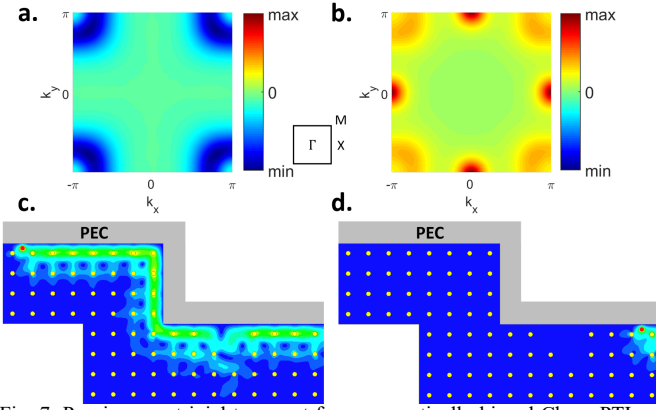

Fig. 7. Proving non-trivial transport for a magnetically biased Chern PTI. a. Berry curvature for the second band, showing large dips near the M point. Middle inset: BZ diagram. b. Berry curvature for the third band, with sharp peaks at the X point and smaller peaks at the M point. c. Example of robust transport, with two 90-degree bends and a defect along the path. Note the lack of backscatter. d. Exciting the same structure from the other end shows the isolator-like behavior, with no mode for propagation in the opposite direction.

$$\vec{\vec{\mu}} = \begin{bmatrix} \mu & i\kappa & 0 \\ -i\kappa & \mu & 0 \\ 0 & 0 & \mu_0 \end{bmatrix}. \tag{21}$$

Here $\kappa$ represents the effect of the $z$-directed DC magnetic field and is zero when the field is turned off. For a 1600 Gauss magnetic field [33], the values at 4.28 GHz are $\kappa = 12.4\mu_0$ and $\mu = 14\mu_0$, with $\mu_0$ being the vacuum permeability in MKS. Breaking TRS by turning on the magnetic field opens a bandgap near the degenerate point for the 2nd and 3rd bands at the M point, as shown in Fig. 6b.

To confirm that the opened bandgap is indeed topologically non-trivial, we can observe the behavior of the Berry curvature for the various bands, shown in Fig. 7 a-b. We can see that the 2nd mode has a very large contribution to the Berry curvature right at the M point, and likewise for the X point of the 3rd mode. Integrating over each separately, we find the Chern number of the lowest band to be zero, while the next three are -1, 2, and 1, indicating the existence of edge modes within the upper bandgaps.

An important thing to note is that while the Chern number $C_n$ is associated with each band $n$ of the bulk, edge modes are associated with the bandgaps *between them*. To differentiate this, we often speak of the "gap Chern number" $C_{gap} = \sum_{n<n_g} C_n$, which is just the sum of the Chern numbers of all bands below a given bandgap with upper band $n_g$ [16]. Hence, to observe edge modes, we need to operate within a bandgap between two materials with different gap Chern numbers, with the *net* number of modes being their difference, $N_{modes} = C_{\Delta gap} = |C_{gap,1} - C_{gap,2}|$. For this example, that implies the lowest bandgap will have no edge modes (or, more precisely, the net number of rightward modes equals the net number of leftward modes [35]), while the second and third bandgaps will.

The magnetized YIG model displays all the tell-tale signs of a Chern PTI, and as such we can construct a wide range of devices that exploit its non-reciprocal and highly robust nature. One such demonstration is an isolating transmission line with two 90-degree bends, shown in Fig. 7c-d (compare with Fig. 1). Similarly to ferrite-based magnetic isolators, the device is non-reciprocal for any EM mode inside the non-trivial bandgaps, but there are a few important and highly attractive features:

1. Unlike a traditional isolator, EM energy is not simply routed to a lumped element load and dissipated locally as heat [34]. Instead, the influence of the Berry phase results in an edge mode only for a single direction of propagation, with no allowed modes in the opposite direction. Hence, any energy sent the opposite way will be either reflected or decay exponentially into the bulk in the same manner as a trivial photonic crystal (Fig. 7d).
2. The directionality of the mode is determined by the direction of the bias magnetic field, so flipping its direction will also flip the allowed propagation direction.
3. With the lack of backward modes, a source of backscatter, like the shown 90-degree bends, will force the energy around corners with negligible losses. This will occur so long as the strength of the scatterer is not greater than the size of the non-trivial bandgap, provided the scatterer is non-magnetic [2]. Likewise, any small defect, like the three missing rods, will not lead to scattering.
4. Being essentially a distributed device, the level of isolation and insertion loss can be tuned by varying the length and shape of the structure.

This example is for a 2D system, which can be experimentally emulated via a parallel plate waveguide structure, with the separation between the plates being very thin, ensuring only TM modes can propagate. This platform makes it easy to analyze but is less straightforward to integrate into normal EM and photonic systems. However, there are numerous studies and ongoing work to create Chern PTIs for more practical settings [36]. A major breakthrough for this line of research was the development of the valley and spin PTIs (detailed in the following sections), which remove the requirement of the external magnetic field.

**Side note:** Although the example of a Chern PTI studied here concerns periodic structures with discrete translational symmetry, it has been recently shown that Chern numbers can also be defined for continuous media, such as a homogenous magnetized plasma [37]. In addition, other platforms including arrays of coupled waveguides can be used to emulate the effect at optical frequencies [38].

## VI.  VALLEY PTIs

Although the Chern PTI has many advantages, the requirement for magneto-optical materials and external magnetic fields places limits on the practical applications. The question then arises: can we construct a PTI with similar features of robustness to disorder or sharp turns, while still being reciprocal? The answer turns out to be yes, with some limitations. There is another type of PTI made of passive materials that exploits an inherent degree of freedom of hexagonal lattices that can be used to mimic similar phenomena for robust edge state propagation, although the level of robustness depends on the types of disorder considered. Specifically, in a hexagonal/graphene-like lattice, the angular rotation of the E fields at the high-symmetry point K or K' generates an intrinsic magnetic moment which is called the "valley degree of freedom" [19]. The term "valley" is used owing to the shape of the dispersion near the K (K') point, which in a triangular lattice is a deep dip or a sharp peak, or valleys. Just as the Chern PTI emulates the quantum Hall effect, the valley PTI is a model for the "quantum valley Hall effect," studied in graphene-like materials.

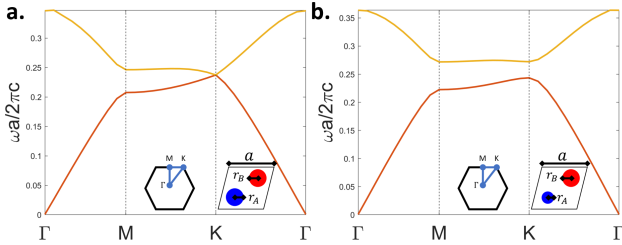

Fig. 8. Band diagram for the a. symmetrical (trivial) and b. asymmetrical (non-trivial) valley structure, with Si ($\epsilon_r = 11.9$) rods in air. Inset left: BZ path, right: unit cell, with a. $r_A = r_B = 0.25a$, b. $r_A = 0.19a, r_B = 0.25a$.

To design a valley PTI, we can start from a graphene-like PhC, which will possess the "Dirac"-like degenerate point at the K (K') point. Such a lattice can be constructed by a unit cell containing two rods of equal radius (A and B sites), shown in the inset to Fig. 8a. Like when constructing a Chern PTI, a symmetry must be broken to lift the degeneracy. In a valley PTI, a controllable bandgap can be achieved by differentiating the A and B rods in the unit cell, thus breaking the inversion symmetry. As we will show through examples, a graphene-like PhC that lacks inversion symmetry exhibits opposite Berry curvature at K and K' points [23]. In principle, this allows us to selectively couple to either K or K' valleys, which would result in a unidirectional topologically protected edge mode that is locked to the direction of Γ to K or Γ to K'.

Here we look at an example of a dielectric valley PTI from [39]. Each unit cell consists of two silicon rods, A and B, with corresponding radii $r_A$ and $r_B$. When $r_A$ and $r_B$ are identical (here $r_A = r_B = 0.25a$, with lattice constant $a$), the structure becomes a type of photonic graphene, and there is a Dirac degeneracy at the K(K') point as shown in Fig. 8a.

We then break the inversion symmetry by shrinking the A rod slightly ($r_A = 0.19a$). This lifts the degeneracy and opens a complete bandgap around it, as shown in Fig. 8b. Note that we can tune this bandgap by tuning A and/or B's dimensions. The more different A and B are, the larger the bandgap.

As shown in Fig. 9, the in-plane E-field distribution of the first and second bands at the K valley are accompanied with an energy flux (i.e. time-averaged Poynting vectors) rotating in either a clockwise or counterclockwise manner. In accordance with TRS, we also find that the field profile at the K' valley exhibits the reversed direction of energy flux. The flux vortex's center corresponds to a singular point of the phase (here of out-of-plane electric field, $E_z$), carrying an orbital-angular-momentum (OAM) with its sign depending on the vortex direction [39]. This vortex can be considered as an "artificial magnetic field"-like effect that replaces the role of the real magnetic field of the Chern PTI. Meanwhile, inverting the orientation of the unit cells in plane (i.e. swapping A and B lattice sites) results in identical band structures but opposite signs of OAM at the K and K' valleys. Importantly, the frequency order of the OAM states at each valley is also inverted, which indicates a topological phase transition.

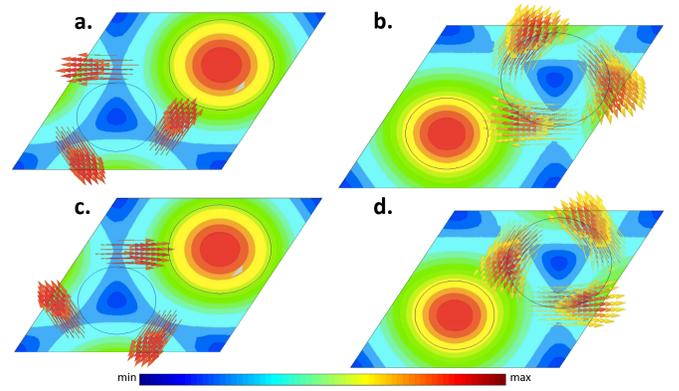

Fig. 9. Illustration of the difference in energy flux of the K and K' modes for the upper and lower bands. Color denotes the magnitude of the E field, while the arrows are the real part of the Poynting vector. a. Lower band at K. b. Upper band at K. c. Lower band at K'. d. Upper band at K'.

To further validate the non-trivial topological character of the bands, we can numerically calculate the Berry curvature, as shown in Fig. 10a-b. The spike at the K point results in a Berry phase of $\pi$, while there is a $-\pi$ Berry phase accumulation at K'. Integrating over half of the BZ (or near K (K')), we get the valley Chern number of $+\frac{1}{2}$ for K and $-\frac{1}{2}$ or K'. If the A and B sites are exchanged, we will get the valley

> **Side note:** How can the "valley Chern number" be $\pm\frac{1}{2}$ when the Chern number is always an integer? The value is only guaranteed to be an integer when the integrated curvature is *closed*, while here we consider only half the BZ, which is therefore open. Valley PTIs gain their characteristics by the opposite behavior at K and K', even though the "true" Chern number is zero.

Chern number of $-\frac{1}{2}$ for K and $+\frac{1}{2}$ or K'.

Furthermore, we can see that the signs of both flip from the lower band to the higher band. This indicates that valley-polarized topological edge states exist within the bandgap at an interface between structures with opposite unit cell orientation (between A-B and B-A). The interface will therefore have A sites adjacent to A sites, or B sites adjacent to B sites. The number of edge mode at each valley, in accordance with the bulk-edge correspondence, is determined by the difference of $C_v$ above and below the bandgap: $C_{v,2}^K - C_{v,1}^K = 1$, $C_{v,2}^{K'} - C_{v,1}^{K'} = -1$, where each value corresponds to a single valley (K or K'). The differing signs here could be interpreted as the sign of the group velocity of the forward and backward propagating modes at the two valleys [19].

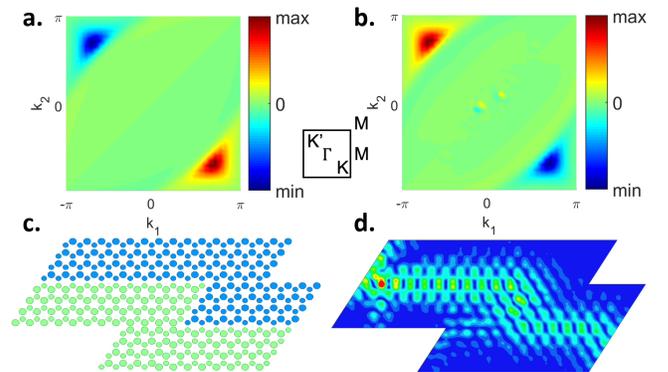

Fig. 10. Proving non-trivial transport of a valley PTI of two-site silicon rods. a. Berry curvature for the lower band. Middle inset: BZ diagram. b. Berry curvature of the upper band, showing flipped peaks and nulls from the lower case. c. Interface structure with two 120-degree bends. The two regions are simply rotated versions of each other. d. Magnitude of the E field of the structure.

We can again build essentially the same bent waveguiding model as the Chern PTI to demonstrate the robustness of the valley structure, shown in Fig. 10c-d. The valley PTIs also have their own features:

1. Unlike the Chern PTIs (and spin PTIs) where topological edge modes can be formed at the interface between a nontrivial lattice and a trivial lattice, the valley edge modes only exist at the interface between two valley PTI with opposite valley Chern number (1/2 and -1/2). Therefore, when constructing a valley waveguide, there must always be a pair of complementary valley PhCs.
2. Since the valley edge modes are coupled to K or K' valleys, these edge modes are the most robust in the directions where the valleys are clearly defined (i.e. Γ to K or Γ to K' directions), indicating that they will preserve their unidirectionality only when sharp turns are 120 degrees. In contrast to Chern PTIs, defects that scatter valleys into each other (breaking the $C_{3v}$ symmetry for example) will weaken the edge mode and give lower robustness [40].
3. Valley states can couple to spin-type modes in certain PTIs [41]. These spin-valley coupled edge states can be used to form "valley splitters" that are not restricted by the orientation of the unit cell [41].
4. Unlike many spin PTI designs, it is simple to construct planar or nearly planar valley PTIs, making them an attractive choice for integration with normal silicon [42] or metal-on-insulator designs.

> **Side note:** It is also possible to build a purely metallic valley PTI. For example, a valley PTI can be constructed by a patch-type flat metasurface to engineer TE modes (or apertures for TM modes) [43]. Simply changing the shape of the patch from a hexagon into a triangle reduces the cell from a 60-degree rotation symmetry to 120-degree rotation symmetry. As a result, the degenerate bands split at the two inequivalent but time-reversed valleys (K and K′), leading to a bandgap in the BZ. Edge modes will form between sheets of upward and downward pointed triangles.

## VII. SPIN PTIs

The last major PTI type, the spin PTI, is in a practical sense similar to the valley type, being reciprocal while still possessing highly robust features. A full example of a recently demonstrated planar metallic PTI [44], which has a number of attractive features for integration with traditional microwave systems [45] is provided in the additional online materials. Spin PTIs are also readily adaptable to dielectric platforms suitable for optical bands. [46]

## VIII. CONCLUSIONS AND FUTURE OUTLOOK

In this tutorial, we have given an overview of the concepts, mathematics, and implementations of photonic topological insulators (PTIs). The central idea that relates the topics together is the geometric phase, which lies at the heart of both theory and physical realization. This concept is readily applied to 2D PhC, which is a simple platform to engineer topological modes. Computations of the Berry curvature, Chern number, and other topological invariants for a given design illuminate how the geometric phase influences the system and can be readily calculated with numerical tools to aid in design.

The three most common formulations of PTIs, the Chern PTI, valley PTI, and spin PTI, all represent different strategies to achieve the effects of topologically protected modes, lending considerable flexibility to their usage. In all cases, such devices possess remarkable robustness to a wide class of disorder, which could enable much greater fabrication tolerances for applications like extremely robust integrated optical waveguides [38]. Likewise, their immunity to backscattering off of sharp bends has the potential to shrink device footprints by eliminating the gradual bends or careful engineering at edges needed to overcome losses or higher order mode mixing when a turn is required in a waveguide [2],[36].

For the case of the Chern PTI devices, there is large effort to deploy topologically protected lasers [47], with many recent studies realizing arbitrarily shaped optical cavities immune to disorder [48]. Further uses can be seen in isolators [34] and circulators [49]. In the magnetic field-free valley and spin implementations, there is potential to use such devices in place of traditional transmission lines or waveguides, with the added benefits of sharp bend immunity and robustness to disorder [44],[40]. For microwave devices, many of these are functionally similar to traditional metallic structures, and as such could be integrated into standard systems with relative ease [40]. At optical frequencies topologically protected designs could enable features like spin-selective filtering [50] and unidirectional polarization control [19], beyond general robustness.